\documentclass{JHEP3}

\title{Poincare field theory for massive particles}

\author{B. Sazdovi\'c \\
        Institute of Physics, P.O.Box 57, 11001 Belgrade, Serbia \\
        E-mail: \email{sazdovic@phy.bg.ac.rs}}

\abstract{

There is  ambitious pretension formulated by  Weinberg  \cite{W} that
{\it any relativistic quantum theory will look at sufficiently low energy  like a quantum field theory.}
It is based on the observation that  for formulation of quantum field theory
{\it ... much better starting point is Wigner's definition of particles as representations of  inhomogeneous Lorentz group} \cite{Wi, BW}.

To achieve that Ref.\cite{W}   starts with particles and get to the field  equations later.
Here we  propose a complementary  approach   and   directly introduce  field  equations  as   Casimir eigenvalue problem.
Note that Casimir invariants    commute with all group elements and therefore commute between each other. So, they have common eigenvalues  (for Poincare group  mass and spin )
and common eigenstates  (here irreducible representation of Poincare group).

We  use derivatives as   standard representation for momenta  $P_a \to i \partial_a$  and
introduce representation for arbitrary spin operator $ S_{a b} $  with the help of   recurrence relations.
To solve eigenvalue problem for Casimir operators we will go to the formulation with standard momentum, where differential equations turn to algebraic ones.
Then for arbitrary field   $\Psi^A $  we construct projection operators on  particular spins.  The   irreducible representations, have a role of equations of motion.

For fermions we can go to linear form of equations of motion  and  obtain Dirac  equation.

}

\usepackage{amsmath}

\begin{document}

\section{Introduction  }
\setcounter{equation}{0}

Quantum field theory \cite{W, BS, Pes}  combines quantum theory   \cite{Dir}   with  principle of special relativity.  In contrast to   traditional approach \cite{BS, Pes}  which start with Maxwell and Dirac equations
we will  prefer   Weinberg  approach   \cite{W}  who uses   Wigner's definition of particles   as irreducible  representations of  Poincare group  \cite{Wi,BW}.

Unlike  Ref.\cite{W}  which starts with particles  and get to the field  equations later we will start with {\it  principle field equations} for  arbitrary  spin
and show that all known equations for free fields follow from our principal  equations.  We offer general   prescription which can be used  to construct   field theory based on  symmetry of some  given group.
Here we apply  general   prescription   to the  Poincare group  which explains  the title  of the present article  {\it Poincare  field theory }.

In this article we start with special theory of relativity as a theory in four dimensional Minkowski space-time with three space  directions $\mathbf{x}$ and one time  direction  $t$.
Poincare transformations are all four dimensional  rotations  (space rotations and boosts)  and translations   which preserve  line element $d x^2 = d t^2 -  d \mathbf{x}^2$.
They are symmetries of Minkowski space-time.
Generators  of these transformations,  translations $P_a$  and  rotations $M_{a b}$,  form a  Poincare  algebra.

\section{Representations of  the Poincare  algebra  }
\setcounter{equation}{0}

We will need  representations of  the Poincare algebra  as a first step  in order to find representations of  the Poincare Casimir invariants.

\subsection{Poincare  group  }

Lie algebra of Poincare group has a form
\begin{eqnarray}\label{eq:PG}
[P_a, P_b] = 0  \, , \qquad  [P_c, M_{a b}]  =   - i \Big( \eta_{b c} P_a  - \eta_{a c } P_b  \Big)   \, ,
\end{eqnarray}
\begin{eqnarray}\label{eq:}
[M_{a b}, M_{c d}]  =  i   \Big( \eta_{a d} M_{b c} + \eta_{b c} M_{a d} - \eta_{a c} M_{b d} - \eta_{b d} M_{a c} \Big)   \,  ,
\end{eqnarray}
where $P_a$ are translation generators and
$ M_{a b}   =   L_{a b} +   S_{a b}$  are four dimensional  rotations generators.  They  consist of orbital part
$L_{a b} =    x_a P_b - x_b P_a $  and spin part $S_{a b}$.

\subsection{Poincare Casimir operators}

Casimir operators  commute  with all  group  generators and  allow us to label the irreducible representations.
Here we will introduce Casimir operators for massive Poincare  group and find theirs representations for arbitrary spins.
In  the case of  $P^2 > 0$  there are   two Casimir operators
\begin{eqnarray}\label{eq:COP}
P^2 = P^a P_a \, ,  \qquad     W^2  =  - \frac{P^2}{2}  M_{a b}  M^{a b}  \, ,
\end{eqnarray}
where
\begin{eqnarray}\label{eq:PLv}
W_a = \frac{1}{2} \varepsilon_{a b c d} M^{b c} P^d     \, ,
\end{eqnarray}
is  Pauli-Lubanski vector.   Note that  Pauli-Lubanski vector does not depend  on orbital part  $L_{a b} $  because
\begin{eqnarray}\label{eq:PLv1L}
 \varepsilon_{a b c d} L^{b c} P^d =   \varepsilon_{a b c d}  (x^b P^c - x^c P^b  )   P^d   =  0  \, .
\end{eqnarray}
Therefore, we can rewrite above equations in the form
\begin{eqnarray}\label{eq:COP1}
P^2 = P^a P_a \, ,  \qquad     W^2  =  - \frac{P^2}{2}  S_{a b}  S^{a b}  \, ,
\end{eqnarray}
where now
\begin{eqnarray}\label{eq:PLv1}
W_a = \frac{1}{2} \varepsilon_{a b c d} S^{b c} P^d     \, ,
\end{eqnarray}
and   $S_{a b}$ is spin part of of Lorentz generators   $M_{a b}$.

Let us stress that   from Casimir  operator's point of view  just  generators $P_a$  and $S_{a b}$   are relevant. They satisfy relations
\begin{eqnarray}\label{eq:PGS}
[P_a, P_b] = 0  \, , \qquad  [P_c, S_{a b}]  =  0  \, ,
\end{eqnarray}
\begin{eqnarray}\label{eq:PGSS}
[S_{a b} , S_{c d}]  =  i   \Big( \eta_{a d} S_{b c} + \eta_{b c} S_{a d} - \eta_{a c} S_{b d} - \eta_{b d} S_{a c} \Big)   \,  .
\end{eqnarray}
Note that generators  $P_a$ and  $S_{a b}$ commute and so  Casimir  operators   are defined  without problem of order  unambiguity.

\subsection{Representation of Casimir operators}

Representations of Poincare group  are labeled by the  eigenvalues of Casimir invariants,   that we can assign to a physical state.
For $P^2 > 0$ eigenvalues of Casimir operators are    mass $m$  and   spin  $s$
\begin{eqnarray}\label{eq:EvCo}
P^2 = m^2 \, ,  \qquad  W^2 =   - m^2   s(s+1) \, .
\end{eqnarray}

To find  representation of Casimir operators we need representation of Poincare algebra  generators,  momentum  $P_a$  and spin   $ S_{a b}$.
Representation of momentum  $P_a$ is well known from quantum mechanics $(P_a)^A{}_B \to i   \delta^A_B  \partial_a$ and it is  spins independent.

We can  obtain   representation of   spin generators  $ (S_{a b})^A{}_B$  for  arbitrary field   from corresponding expressions with smaller spins and initial expression for fermions.
Starting from infinitesimal Lorentz  transformation
\begin{eqnarray}\label{eq:iLt}
\Psi_\omega^A (x)  = \Psi^A (x) - \frac{i}{2} \omega^{a b}  (S_{a b})^A{}_B  \Psi^B (x)    \, ,
\end{eqnarray}
we can  find    infinitesimal  transformation of product $\Psi_1^A \Psi_2^B$. First,  multiplying expressions   (\ref{eq:iLt})  for  infinitesimal  $ \omega^{a b}$  we obtain
\begin{eqnarray}\label{eq:}
\Psi^A_{1 \omega} (x) \Psi^B_{2 \omega} (x)
 = \Psi^A_1 (x)\Psi^B_2 (x)  - \frac{i}{2} \omega^{a b}   \Big[  (S_{a b})^A{}_C  \Psi^C_1 (x) \Psi^B_2 (x)  +  \Psi^A_1 (x) (S_{a b})^B{}_D  \Psi^D_2 (x)   \Big]   \, .  \qquad
\end{eqnarray}
Second,   by definition of Lorentz  transformation  we have
\begin{eqnarray}\label{eq:}
\Psi^A_{1 \omega} (x) \Psi^B_{2 \omega} (x)   =  \Psi^A_1 (x)\Psi^B_2 (x)  - \frac{i}{2} \omega^{a b}  (S_{a b})^{A B}{}_{C D} \Big( \Psi_1^C (x) \Psi_2^D (x) \Big)  \, .
\end{eqnarray}
Comparing these results we can conclude that  spin generators   $(S_{a b})^A{}_B$ act as derivatives, where its form depend on the fields on the right
\begin{eqnarray}\label{eq:BEder}
 (S_{a b})^{A B}{}_{C D}    =  (S_{a b})^{A}{}_{C} \delta_D^B  +  \delta^A_C (S_{a b})^{B}{}_{D}  \, .
\end{eqnarray}

It is easy to check that
\begin{eqnarray}\label{eq:}
 [ S_{a b},  S_{c d} ]^{A B}{}_{C D}
=    [ S_{a b},  S_{c d} ]^A{}_C     \delta_D^B  +  \delta^A_C  [ S_{a b},  S_{c d} ]^B{}_D     \, ,
\end{eqnarray}
so that $(S_{a b})^{A B}{}_{C D} $  is solution of  (\ref{eq:PGSS})  because $(S_{a b})^A{}_B  $ is solution of   this  relation.

The  initial expression for   Dirac spinor is
\begin{eqnarray}\label{eq:gLr21}
(S_{a b})^\alpha{}_\beta   = \frac{i}{4} [\gamma_a, \gamma_b ]^\alpha{}_\beta  \, ,
\end{eqnarray}
and  we  can  find  representations  for all other fields  using recurrence relation (\ref{eq:BEder}).
For example, using expression for vectors in terms of spinors  $ V^a (x) =  {\bar \psi} (x)  \gamma^a   \psi (x) $    we can find  the same expression for vector  spin generators as that   obtained by direct calculation
\begin{eqnarray}\label{eq:SOvf}
 (S_{a b})^c{}_d  =  i \Big(    \delta^c_a   \eta_{b d} - \delta^c_b   \eta_{a d}  \Big)     \, .
\end{eqnarray}

\section{Principle field equations}
\setcounter{equation}{0}

The fields are functions defined in Minkowski space-time  which transforms with some representation of Poincare group. They  can be marked with arbitrary number of  vector and spinor indices. We will
introduce notation $\Psi^A (x)$ where $A$ contains the set of corresponding indices.

In order to separate states   which describe  definite  particles we should impose some constraints on the field $\Psi^A (x)$.
Since in field theory  particles  are defined by mass and spin it is natural to use just operators whose eigenvalues  are mass and spin.
Consequently, {\it we will postulate  principle field equations for  arbitrary  spin as   representation   of  relations  (\ref{eq:EvCo}) }
\begin{eqnarray}\label{eq:BS}
(P^2)^A{}_B  \Psi^B (x)  =  m   \Psi^A (x)  \, ,  \qquad   {\cal S}^A{}_B    \Psi^B (x) =  s (s + 1)  \Psi^A (x)   \, .
\end{eqnarray}

In fact they are   Casimir eigenvalue equations.  These equations  are  Poincare covariant because Casimir operators commute with all Poincare generators.
In particular,  Casimir operators  commute mutually.  Since  commuting  observables  have  a complete set of common eigenfunctions we are able to impose both   Casimir operators to the same field $\Psi^A (x)$.

In our basic equations   (\ref{eq:BS})  $(P^2)^A{}_B$ and ${\cal S}^A{}_B $ are representation of Casimir operators.
The eigenvalues  of the operators   $(P^2)^A{}_B$ and    ${\cal S}^A{}_B$  are $m$ and   $s (s + 1)$ and they define the mass $m$  and  the  spins $s$. The eigenfunctions $\Psi^A (x)$    are irreducible representations of the Poincare group.   So  irreducible representations are  labeled by the  mass $m$ and the  spin  $s$.

Using representations for  momentum and spin operators  we obtain the  set of differential  equations
\begin{eqnarray}\label{eq:BS1}
\Big( \partial^2 + m^2 \Big)  \Psi^A (x)  = 0  \, ,  \qquad {\cal S}^A{}_B    \Psi^B (x) =  s (s + 1)  \Psi^A (x)   \, ,
\end{eqnarray}
where
\begin{eqnarray}\label{eq:Oper}
{\cal S}^A{}_B  \equiv - \frac{(W^2)^A{}_B}{m^2} =  - \frac{1}{m^2}     (S^a{}_c)^A{}_C  (S^{c b})^C{}_B \partial_a \partial_b   +  \frac{1}{2}   (S_{a b})^A{}_C  (S^{a b})^C{}_B     \, .
\end{eqnarray}

\subsection{Principle  field equations  for standard momentum}

So far we have  obtained  constraints on the fields $\Psi^A (x)$ as eigenfunctions of Casimir invariants.
The next step is to find   projectors on irreducible representations of Poincare group,  which correspond to  well known field  equations.
In order to  to achieve that  it is useful to go to standard momentum.

Lorentz invariant functions of momentum $p^a$  are square    $p^2 = \eta_{a b} p^a p^b$ and sign of $p^0$.  So, for  $p^2$   and sign of $p^0$ defined  in advance we can chose {\it standard momentum }  $k^a$ and
express any momentum  $p^a$  as Lorentz transformation of $k^a$
\begin{eqnarray}\label{eq:paLka}
p^a =  L^a{}_b  (p)  k^b \, .
\end{eqnarray}
For massive case where  $p^2 = m^2$  we can chose   rest frame  momentum   $k^a = (m, 0, 0, 0  )$  as standard momentum.
With this choice the first equation  (\ref{eq:BS1})  is solved.

For standard momentum second  differential equation (\ref{eq:BS1})   becomes  algebraic one, which makes the calculation much easier.
After solving  algebraic equation we can go back to  $p^a$ dependent solutions using  (\ref{eq:paLka}) and then to solution in coordinate representation.

The spin  equation for  standard momentum   takes a form
\begin{eqnarray}\label{eq:BSms}
{\cal S}^A{}_B    \Psi^B (k) =  s (s + 1)  \Psi^A (k)   \, ,
\end{eqnarray}
where
\begin{eqnarray}\label{eq:Operr1}
{\cal S}^A{}_B  =     (S^2_i)^A{}_B    \, ,  \qquad  (S_i)^A{}_B    = \frac{1}{2} \, \varepsilon_{i j k} (S_{j k} )^A{}_B  \, .
\end{eqnarray}

Note that instead of  $6$ components of spin operator $S_{a b}$ in the frame of standard  momentum   we have left with $3$  components  $S_i$,  which are generators of space rotations.
They form a subgroup known as {\it little group} for massive Poinacare case.

The next step is to  solve the eigenvalue problem of operator  ${\cal S}^A{}_B$  (\ref{eq:BSms}).
To have nontrivial solution for function $\Psi^A $ the    characteristic   polynomial   must vanish
\begin{eqnarray}\label{eq:DetSs}
  \det \Big(  {\cal S}^A{}_B  -   \lambda \delta^A_B   \Big)  = 0      \, ,         \qquad   \lambda   \equiv   s ( s + 1)  \, .
\end{eqnarray}
Generally, it  may have multiple solutions, $\lambda_i$  where  $i = 1, 2, \cdots , n$  counts irreducible  representations.
The values $s_i$,  corresponding  to the  eigenvalues $\lambda_i$,   are spins of   irreducible  representations.

The   representations  of eigenfunctions  $\Psi_i^A $  with definite spin   have a form
 \begin{eqnarray}\label{eq:IrrPa}
\Psi_i^A  = ( P_i )^A{}_B \Psi^B   \, ,     \qquad
( P_i )^A{}_B =   \frac{  \Big[  \prod_{j \neq i}^n    \Big(  {\cal S}   - \lambda_j    \Big) \Big]^A{}_B }{  \prod_{j \neq i}^n    \Big(  \lambda_i   - \lambda_j  \Big) }        \, ,  \qquad
 i =  \{ 1, 2, \cdots , n  \} \, ,
\end{eqnarray}
where $(P_i)^A{}_B$ are corresponding  projection operators.
In the case of   degeneracy, we should chose basis   in subspace  $\Xi_\lambda $   of  eigenvalue $\lambda$.
The  fields $\Psi_i^A  (k)$   are representations of Poincare group,   but not necessary  irreducible representations. To  obtain   irreducible representations we should  separate fields
$\Psi_i^A  (k)$ in several  sets  with well-defined symmetry properties.

According to   (\ref{eq:IrrPa})  we will  introduce  the equation of motion for fixed spin  $s_i$
\begin{eqnarray}\label{eq:IrrPa1}
( P_i )^A{}_B \Psi^B  = 0  \, ,     \qquad   i =  \{ 1, 2, \cdots , n  \} \, .
\end{eqnarray}

We can rewrite equation   (\ref{eq:DetSs})   in the form
\begin{eqnarray}\label{eq:Smlade}
\det \Big(  {\cal S}^A{}_B  -   \lambda \delta^A_B   \Big)
=     \prod_{i =0}^n   ( \lambda_i  - \lambda )^{d_i}  = 0         \,   ,    \qquad
\end{eqnarray}
where  $d_i$ are  dimensions of eigenpaces which correspond   to projector  $P_i$.
Note that    $d_i$  is  number of degrees of freedom of component   $\Psi_i^A  (k)$.

The gauge transformations of field  equation  are generated by projection operators. Fields of  different spins  have different gauge transformations. Explicitly
\begin{eqnarray}\label{eq:Gautr}
  \delta   \Psi^A (x)  =  \sum_{j \ne i} (P_j)^A{}_B \, \varepsilon^B_j  \,  ,  \qquad
\end{eqnarray}
is symmetry transformation of field equation   (\ref{eq:IrrPa1})   because projectors are orthogonal. Here, $\varepsilon^B_j$ for  $j \ne i$ are parameters of gauge transformation.

\section{Examples  }
\setcounter{equation}{0}

We are going to  confirm  that above principle field equations for particular spins coincides with  well known field equations.  Here   we will present   three  examples: Klein-Gordon   equation for scalar fields  ($s = 0$),
Dirac  equation for  spinors  ($s = \frac{1}{2}$) and    equation for massive  vector fields  ($s_0 = 0$  and  $s_1 = 1$).

\subsection{ Scalar  field   }

A scalar field has no indices  $\Psi^A (x) \to  \varphi  (x)$,   so that  $(S_{a b})^A{}_B  \to  0 $.  Equation   (\ref{eq:DetSs})  produces  $\lambda= 0$   and consequently  $s = 0$.
This is one dimensional representation with one degree of freedom.
We are left with the  Klein-Gordon   equation
\begin{eqnarray}\label{eq:}
\Big( \partial^2 + m^2 \Big)  \varphi (x)  = 0 \, .
\end{eqnarray}

\subsection{ Dirac  field   }

For   Dirac  field   $\Psi^A (x) \to \psi^\alpha (x)$ and  representation of  spin operator  is
\begin{eqnarray}\label{eq:gLr2f}
(S_{a b})^A{}_B   \to  (S_{a b})^\alpha{}_\beta  =  \frac{i}{4} [\gamma_a, \gamma_b ]^\alpha{}_\beta     \, ,
\end{eqnarray}
so that
\begin{eqnarray}\label{eq:gLr2f1}
S_i   = \frac{1}{2} \, \varepsilon_{i j k} S_{j k}    = \frac{i}{4}  \varepsilon_{i j k}  \gamma_j  \gamma_k   \, ,  \qquad
{\cal S}^\alpha{}_\beta  =   [ (S_i)^2 ]^\alpha{}_\beta   =  \frac{3}{4 }  \delta^\alpha_\beta        \, .
\end{eqnarray}

Then the spin equation  (\ref{eq:BSms})  takes the form
\begin{eqnarray}\label{eq:BSs1d}
  {\cal S}^\alpha{}_\beta      \psi^\beta  (k) =  \lambda   \psi^\alpha  (k)   \, ,  \qquad   \lambda  \equiv  s (s + 1)
\end{eqnarray}
and since $ {\cal S}^\alpha{}_\beta$ is diagonal we obtain
\begin{eqnarray}\label{eq:}
\det (  {\cal S} - \lambda  )^\alpha{}_\beta
 =   \Big( \lambda -  \frac{3}{4}  \Big)^4 =     0             \,  .
\end{eqnarray}
Therefore,  $\lambda  =  \frac{3 }{4}$  which produces spin  $ s = \frac{1 }{2}$.  According to exponent  in last expression this is four dimensional representation,
which means that field $\psi^\alpha  (k) $  has four degrees of freedom.

There is only one trivial projection operator $P^\alpha{}_\beta  =   \delta^\alpha_\beta$.
Using   (\ref{eq:paLka})   we can go from standard momentum  $k^a$  to arbitrary one $p^a$  and then to coordinate representation.  So, we obtain standard
Klein-Gordon equation for all components $(\partial^2 + m^2  )  \psi^\alpha  (x) =0 $.

We can linearize  it    in the form
\begin{eqnarray}\label{eq:DiE}
(i\gamma^a \partial_a + m ) \psi^\alpha (x) = 0 \, ,
\end{eqnarray}
where $\gamma^a$ are constant matrices. In fact,  it produces  Klein-Gordon  equation if
\begin{eqnarray}\label{eq:Gam}
\{\gamma^a, \gamma^b\}  = 2 \eta^{a b} \, .
\end{eqnarray}

Therefore, for fields with  spin $s = \frac{1 }{2}$  in expression  (\ref{eq:DiE}) we   recognize  Dirac equation and in  (\ref{eq:Gam}) relation for $\gamma$- matrices.

\subsection{ Vector  field   }

For  vector field   we have    $\Psi^A  \to V^a $, and  from representation of spin operators   (\ref{eq:SOvf})  we obtain
\begin{eqnarray}\label{eq:}
(S_i)^a{}_b   =   \frac{1}{2} \, \varepsilon_{i j k} (S_{j k})^a{}_b  = i \, \varepsilon_{i j k}   \delta^a_j \, \eta_{k b}      \, ,   \qquad
{\cal S}^a{}_b =   (S^2_i)^a{}_b  =  - 2  \delta_{j k}     \delta^a_j \, \eta_{k b}    \, .
\end{eqnarray}
Therefore, the  spin equation in the rest frame takes the  form
\begin{eqnarray}\label{eq:COPfvr}
{\cal S}^a{}_b     V^b (k)  =  \lambda  V^a   (k )  \, .   \qquad   \lambda   \equiv s (s + 1)
\end{eqnarray}
The   consistency condition   produces
\begin{eqnarray}\label{eq:}
\det (  {\cal S} - \lambda  )^a{}_b = - \lambda  ( 2  - \lambda )^3  = 0      \,  ,
\end{eqnarray}
with solutions for eigenvalues
$\lambda_0    = 0 \, ,  \,     \lambda_1    = 2 $     and for spins    $ s_0 = 0 \, , \,   s_1 = 1$.
Subspace with $ s_0 = 0$ is one dimensional  while one with  $s_1 = 1$ is three dimensional.

Using expression   (\ref{eq:IrrPa})   we obtain two projectors
\begin{eqnarray}\label{eq:PiAB2}
( P_0 )^a{}_b (k)  =  \delta^a_b    -  \frac{  {\cal S}^a{}_b   }{   2   }   =   \delta^a_0  \delta^0_b         \, ,  \qquad
( P_1 )^a{}_b  (k)  =  \frac{  {\cal S}^a{}_b   }{   2   } =  \delta^a_b   -  \delta^a_0  \delta^0_b         \, .
\end{eqnarray}

In order to obtain irreducible representations of vector fields in arbitrary frame we should boost corresponding equation of the rest frame. Then for projection operators we get
\begin{eqnarray}\label{eq:PRdefb3}
(P^L)^a{}_b  (p)    =  {\cal B}^a{}_c  (P_0)^c{}_d   (k)    ({\cal B}^{-1})^d{}_b   =  {\cal B}^a{}_0      ({\cal B}^{-1})^0{}_b   \, ,   \nonumber \\
(P^T)^a{}_b  (p)   =  {\cal B}^a{}_c  (P_1)^c{}_d   (k)   ({\cal B}^{-1})^d{}_b     = \delta^a_b -  (P^L)^a{}_b  (p)    \, .
\end{eqnarray}
 Using expressions for boost components
\begin{eqnarray}\label{eq:}
   {\cal B}^a{}_0  =  \frac{p^a}{m}    \, ,   \qquad     ({\cal B}^{-1})^0{}_a   =    \frac{p_a}{m}    \, ,
\end{eqnarray}
and  condition  $p^2 = m^2$   we obtain  standard form of longitudinal and transversal projection operators
\begin{eqnarray}\label{eq:PRdefbF}
(P^L)^a{}_b  (p)     =  \frac{p^a p_b}{p^2}      \, ,   \qquad
(P^T)^a{}_b  (p)   =  \delta^a_b -  \frac{p^a p_b}{p^2}      \, .
\end{eqnarray}

Therefore,  solution of the spin equation   produces  two irreducible representations
\begin{eqnarray}\label{eq:IRvf}
 s = 0 :   \qquad  V_L^a (p) = (P^L)^a{}_b  (p) V^b (p) =  \frac{p^a p_b}{p^2}   V^b (p) \, , \qquad  \qquad   \nonumber \\
 s = 1  :   \qquad    V_T^a (p) = (P^T)^a{}_b  (p) V^b (p) =  V^a (p) - \frac{p^a p_b}{p^2}   V^b (p) \, ,
\end{eqnarray}
and in coordinate space using $p_a = i \partial_a$
\begin{eqnarray}\label{eq:IRvfc}
 s = 0  :  \quad  V_L^a (x) =  \frac{\partial^a \partial_b}{\partial^2}   V^b (x) \, ,   \qquad
 s = 1 :   \quad    V_T^a (x)  =  V^a (x) -   V_L^a (x)   \, .
\end{eqnarray}

Equations of motion for massive vector fields,  for spins $0$ and $1$  have a form
\begin{eqnarray}\label{eq:SIRcv}
\Big( \partial^2 +  m^2 \Big)  V_L^a (x)  = 0   \, , \qquad     \Big( \partial^2 +  m^2 \Big)  V_T^a (x)  = 0         \, .
\end{eqnarray}
Since the number of degrees of freedom is equal to  dimensions   of representations field $V_L^a (x)$  has  one degree of freedom   and   field $V_T^a (x)$
has   three  degrees of freedom.

If we   introduce notation $\varphi = \partial_b V^b$  so that
\begin{eqnarray}\label{eq:PrO1m}
  V^L_a =    \frac{ \partial_a}{\partial^2} \varphi  \, ,  \qquad                     V^T_a =  U_a                \, ,
\end{eqnarray}
we can  rewrite above equations  as
\begin{eqnarray}\label{eq:COP2vaLf}
\Big( \partial^2 + m^2 \Big) \varphi  (x)  = 0    \, ,  \qquad \Big( \partial^2 + m^2 \Big) U_a  (x)  = 0 \, ,
\end{eqnarray}
where  according to second relation    (\ref{eq:IRvf})  $U_a$ satisfy condition $\partial_a U^a = 0 $.  This condition reduces four  components of the field $U_a$ to three degrees of freedom.  It provides positivity of vector field energy \cite{BS}.

The gauge transformations for spins  $s = 0$  and $s = 1$ take a form
\begin{eqnarray}\label{eq:GTVtl}
 \delta   V^a (x)  =   (P_T)^a{}_b  \varepsilon^b  \,  ,  \qquad   \delta   V^a (x)  =   (P_L)^a{}_b  \varepsilon^b  \,  .
\end{eqnarray}
which for scalar field means $\delta  \varphi =  0$. In fact, first gauge transformation (\ref{eq:GTVtl})  reduces four components of the field $V^a$ to one degree of freedom.

\section{Conclusion }
\setcounter{equation}{0}

We formulated general  approach   which allows us to  obtain free field equations only on the basis of symmetry group.
In order to realize it we need to know  Casimir invariants of  the  symmetry group.  Our approach  shows that symmetry group completely determine corresponding free field theory.

In the present article we started with well known fact that Poincare invariance is the fundamental symmetry in particle physics.  Based on Poincare  Casimir invariants,
we introduced principle field equations  which produce  all known  equations for free  massive  fields  with  arbitrary spin.
In the present article we   considered  massive case and in the next one  \cite{S1}  we will consider massless case.

Therefore, we have developed systematic  approach  that supports   Wigner's  idea  \cite{Wi,BW}   developed  by  Weinberg   \cite{W}  that
particles are    irreducible  representations of  Poincare group.

\end{document}